\documentclass[a4paper,twocolumn]{article}
\usepackage[top=2.4cm, bottom=2.5cm, left=2cm, right=2cm]{geometry}
\usepackage{natbib}
\usepackage{aas_macros}
\usepackage{amsmath}
\usepackage{amssymb}
\usepackage{url}
\usepackage{multirow}

\usepackage{graphicx}

\usepackage{times}
\bibpunct{(}{)}{;}{a}{}{,}
\begin{document}


\title{\flushleft \textbf{Refined position angle measurements for galaxies of the SDSS Stripe 82 co-added dataset}}

\author{J\'ozsef Varga$^{1}$\footnotemark, Istv\'an Csabai$^{1}$ and L\'aszl\'o Dobos$^{1}$}


\maketitle

 
\abstract{Position angle measurements of Sloan Digital Sky Survey (SDSS) galaxies, as measured by the surface brightness profile fitting code of the SDSS photometric pipeline \citep{Lupton 2001}, are known to be strongly biased, especially in the case of almost face-on and highly inclined galaxies. To address this issue we developed a reliable algorithm which determines position angles by means of isophote fitting. In this paper we present our algorithm and a catalogue of position angles for 26397~SDSS~galaxies taken from the deep co-added Stripe~82 (equatorial stripe) images. Data are published on-line at http://www.vo.elte.hu/galmorph.}

\renewcommand{\thefootnote}{\fnsymbol{footnote}}
\footnotetext[1]{E-mail: jozsef-varga@caesar.elte.hu}
\renewcommand{\thefootnote}{\arabic{footnote}}
\footnotetext[1]{Department of Physics of Complex Systems, E\"{o}tv\"{o}s Lor\'and University, Pf. 32, H-1518 Budapest, Hungary}


\section{Introduction}

Reliable shape and position angle measurements of galaxies are essential for a number of fields of extragalactic astronomy including the investigation of correlations between the orientation and spatial distribution of galaxies, weak lensing, etc. Shape measurement algorithms, however, are known to be subject to various unwanted effects (e.g. \citealp{Byun 1995}). Assuming that the pixels of a galaxy are distributed elliptically -- which is a rather simple yet suitable approximation in most cases -- two primary parameters are to be obtained: the axis ratio $b/a$ and the position angle $\phi$, measured with respect to the north direction. In case of disc galaxies, the axis ratio can be easily transformed into the inclination angle ($i$), hence the direction of the rotation axis can be obtained. There remains an ambiguity in the sign of the rotation axis, however, because the ``approaching'' and ``receding'' sides of the disc are hard to be distinguished based on optical images. Spectroscopic and HI measurements, which would yield the required information, are not yet feasible at the scale of optical surveys. In case of elliptical galaxies, even if they are rotationally supported, the determination of the directions of axes depends even more on the projection effects.

\subsection{Position angle measurements in the SDSS}

The Sloan Digital Sky Survey \citep{York 2000} is the largest wide-field photographic survey to date with more than 400 million catalogued objects. The SDSS catalogue provides extensive information on the morphology of the sources. Ellipticities and position angles are measured using four different methods:

\begin{itemize}

\item The first method is based on surface brightness profile fits. Two models are fitted to the two-dimensional images of each object: a pure exponential and a pure de Vaucouleurs' profile. The model fits give estimates on the axis ratio $(b/a)_{\text{exp}}$ and $(b/a)_{\text{deV}}$, and the position angle $\phi_{\text{exp}}$ and $\phi_{\text{deV}}$. These fits also take the PSF into account.

\item The second method is based on flux-weighted second moments, referred as ``Stokes parameters'' in the catalogue (denoted with $Q$ and $U$). Stokes parameters can be used to reconstruct the axis ratio and the position angle. The performance of these parameters, however, is not ideal at low signal-to-noise ratio. \citep{Stoughton 2002}

\begin{figure*}
\includegraphics{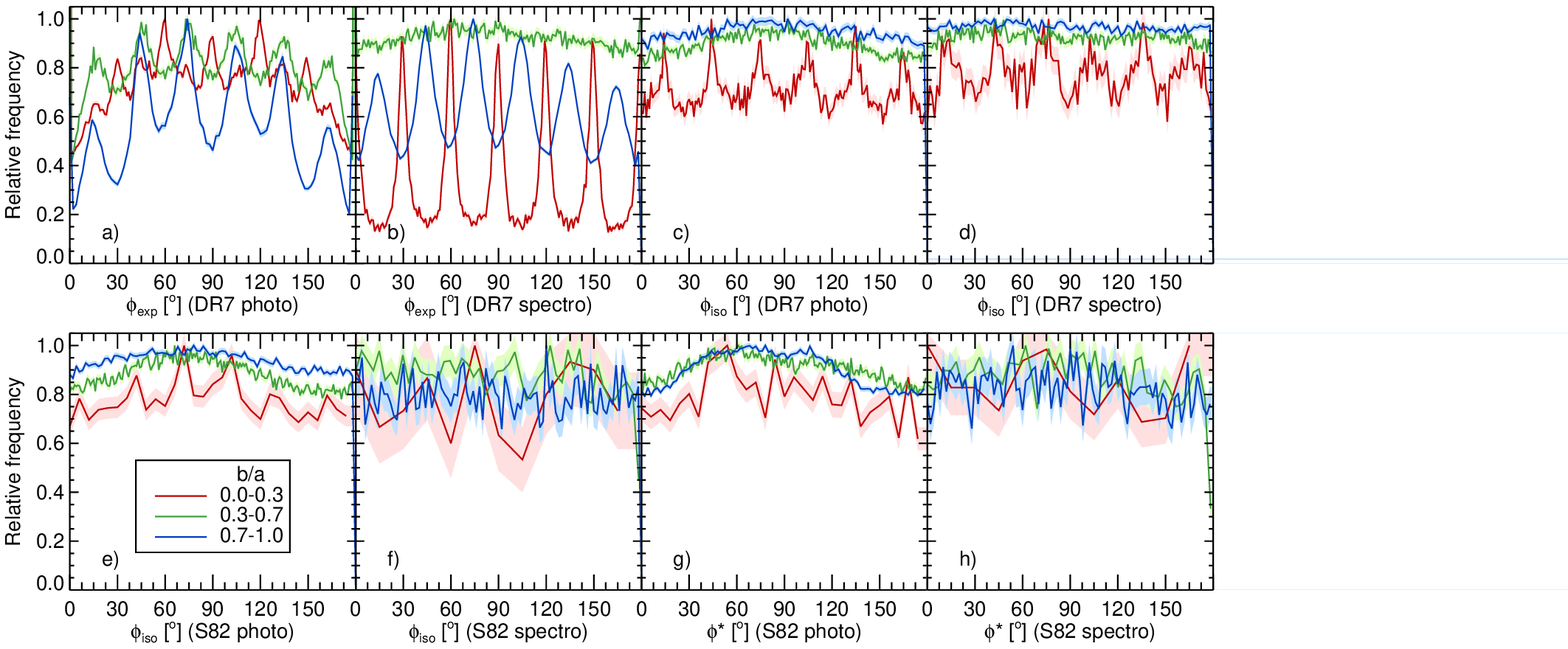}
  \caption{\small Distribution of position angles ($r$ band, respect to the longer image side) of SDSS galaxies. Each panel shows the distribution of the sample divided into three axial ratio bins. Data were collected from the full DR7 catalogue (upper panels), and from the Stripe 82 co-added catalogue (panels e and f). Data in panels g and h are from the current analysis. }
\label{fig:phi}
\end{figure*}
	
\item The third method is based on adaptive moments \citep{Bernstein 2002} which are the second moments of the object intensity measured using a radial weight function iteratively adapted to the shape (ellipticity) and size of the object. This is well suited for cosmic shear measurements, however, contrary to the second method, it is not model independent. \citep{Bartelmann 2012}

\item The last method determines the ellipticity of the $25$~mag~arcsec$^{-2}$ isophote by measuring the radius of the isophote as a function of angle $r(\phi)$. The major and minor axes ($a_\text{iso}$, $b_\text{iso}$) and the position angle ($\phi_\text{iso}$) are then extracted from the coefficients of the Fourier expansion of $r(\phi)$. Isophotal quantities turned out to be unreliable and have been dropped from the catalogue since Data~Release~8 \citep{Aihara 2011}.
\end{itemize}

\begin{table}
\caption{\small Statistics of the source list (number of processed objects and the chosen 
magnitude limits), and the percentage of the fitted and rejected objects in each photometric
filter.}
\label{tab:obj_lst}
\begin{tabular}{ccccc}
\\
\hline
& \multicolumn{2}{c}{Source list statistics} & \multicolumn{2}{c}{Fit statistics} \\
Filter & Number & Mag. limit & Rejected & Fitted \\ 
& & & (\%) & (\%) \\
\hline
u & ~~~80472 & 20 & 53.6 & 46.4\\
g & 1114774 & 22 & 33.9 & 66.1\\
r & 2645180 & 22 & 53.9 & 46.1\\
i & 1956496 & 21 & 41.1 & 58.9\\
z & 1079363 & 20 & 52.9 & 47.1\\
\hline
\end{tabular}
\end{table}

Note that the last three methods do not account for the varying seeing. For more details on SDSS image processing algorithm, refer to \citet{Stoughton 2002}.

\subsection{Biased distribution of SDSS position angles}
\label{sec:bias}

\begin{figure}
\includegraphics{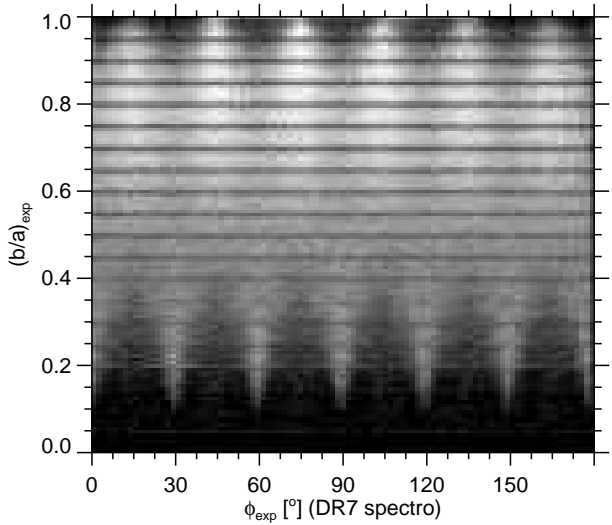} 
  \caption{\small Two dimensional histogram of $r$-band position angles (respect to the longer image side) and axis ratios of galaxies from the spectroscopically confirmed main galaxy sample (same data as in Fig. \ref{fig:phi} panel~b). The square root of the frequency counts is scaled to 8-bit greyscale pixels.}
\label{fig:hist2d}
\end{figure}

In Fig.~\ref{fig:phi} we plot the histogram of position angles from the exponential model fits $\phi_\text{exp}$ (panels~a~and~b), from the isophotal methods $\phi_\text{iso}$ (panels~c~through~f), and from the present analysis $\phi^*$ (panels~g~and~h) for the entire (morphologically identified) galaxy sample (panels~a,~c,~e~and~g) and only for the spectroscopically confirmed main galaxy sample (panels~b,~d,~f~and~h) \citep{Strauss 2002}. The colours refer to three different axis ratio intervals, see figure caption. It is obvious from Fig.~\ref{fig:phi} that the distribution of $\phi_\text{exp}$, as measured by the SDSS pipeline, is heavily biased by systematic effects. A huge periodic modulation can be clearly seen in the relative frequency of $\phi_\text{exp}$ with a period of $30^{\circ}$ in the $b/a = 0.0$-$0.3$ and (with opposite phase) in the $b/a = 0.7$-$1.0$ subsamples (panels~a~and~b). Significant differences between the entire photo sample and the spectro sample are also apparent. While the bias of $\phi_\text{exp}$ is even stronger for edge-on and face-on galaxies in the spectro sample, it disappears for the intermediate $b/a = 0.3$-$0.7$ subsample. Interestingly, $\phi_\text{iso}$ shows similar bias, but only in the case of highly inclined galaxies (red line in panels~c~and~d). We mention that a very similar systematic distortion of $b/a$ is also observable, with a period of 0.1 (see Fig.~\ref{fig:hist2d}). Distributions are plotted for SDSS~DR7 but the bias is still present in DR8 and DR9.

\subsection{Origin of the strong systematic bias}

To our best understanding the strong bias in the distribution of $\phi_\text{exp}$ is caused by the surface brightness fitting algorithm of the SDSS photo pipeline. As fitting a three parameter ($r_e$, $a/b$, $\phi_\text{exp}$) profile to two dimensional images is computationally expensive, the photo pipeline uses precomputed tables of models. To use these tables, first radial profiles have to be binned into $30^{\circ}$ sectors of annuli around the object centroids which introduces a preference of multiples of $30^{\circ}$ in $\phi_\text{exp}$. Apparently, the $\chi^2$-minimization algorithm fails to move away from the local minima of the precomputed models by varying $\phi$ and the $30^{\circ}$ binning propagates into the final fits. The situation is even more severe in case of spectroscopically confirmed galaxies.


\section{Improved isophote fitting algorithm}
\label{sec:algorithm}

Because the available position angle data of SDSS galaxies are unreliable we decided to develop our own isophote fitting algorithm and software tools written in IDL\footnote{The code is available from the authors on request.}. The algorithm automatically identifies the most reliable isophote surface brightness for each frame, determines this isophote of each object and fits an ellipse to it. The main steps of image processing are as follows.

\begin{itemize}
	\item Read in a calibrated image.
	\item Choose a global surface brightness level to find the most reliable isophotes. Here we	fit a Gaussian to the histogram of pixel values of the whole frame, from which we get the mean background level and the noise $\sigma$. The $4\sigma$ level is chosen as the surface brightness of the isophotes.
	\item Detect the isophotes with the CONTOUR routine of the standard IDL library. Isophotes are returned as contour polygons given in pixel coordinates.
	\item Match the objects from the source list with contour polygons. Here we use a matching radius of 3.6'' (9~px).  
	\item Check the ``quality'' of the contours and reject those that are too small to fit reliably, i.e. having less than 10 edges.
	\item Fit ellipses to the contours. Fitting is performed using the MPFIT routine. MPFIT is a port to IDL of the non-linear least squares fitting program MINPACK-1 \citep{Markwardt 2009} and, thus is not part of the standard IDL library. The routine returns five parameters: the $X$ and $Y$ coordinates of the centroid, length of the minor and major axis, and the position angle with respect to the pixel coordinate system.
\end{itemize}

\begin{figure}
\includegraphics{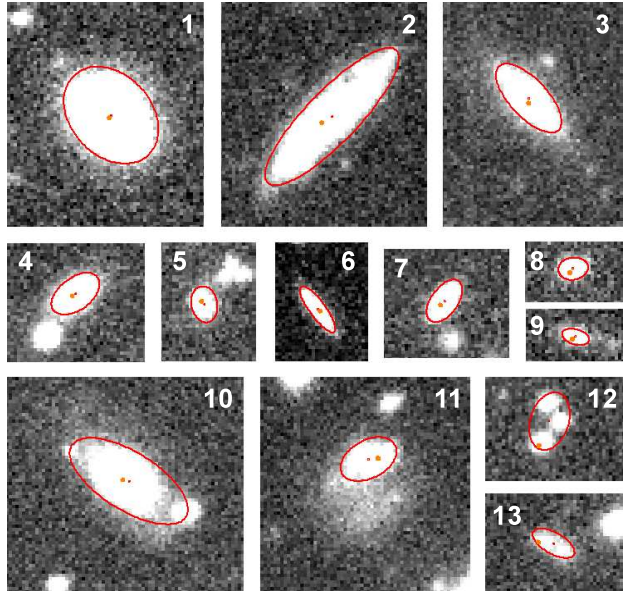}
  \caption{\small Examples of galaxies overplotted with the ellipses fitted to the $4\sigma$ isophotes by our algorithm. Images 1--9 are good fits, while images 10--13 are not. The galaxy in image 11 has complex morphology, hence our elliptical model is not appropriate.
In image 10, 12 and 13 there is an other source so close to the object in question, that
the isophotes are blended, biasing the elliptical fit.}
\label{fig:fit}
\end{figure}

In Fig. \ref{fig:fit} we show several images of galaxies overplotted with the fitted ellipses. See caption for detailed description. 

Fits are automatically rejected if the contours do not match unambiguously a single object or because the contour polygons are too small. A small fraction of the objects is fitted incorrectly due to blended isophotes. The centroids of the fitted ellipses in these cases are a few pixels off the source positions as measured by the SDSS photo pipeline. The percentage of rejected and successfully fitted objects is listed in Tab~\ref{tab:obj_lst}. Depending on the imaging filter a 46--66\% of the objects are fitted successfully.

\begin{figure*}
\includegraphics{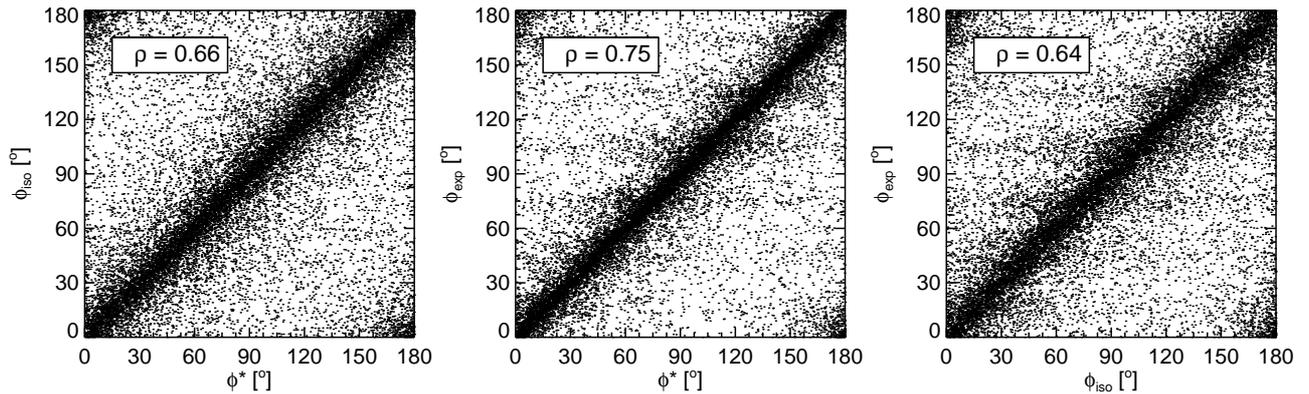}
  \caption{\small Comparison plots of position angles determined by our algorithm ($\phi^*$) and those by the SDSS photo pipeline ($\phi_\text{exp}$ and $\phi_\text{iso}$) for the $r$-band Stripe~82 co-added images. Over 20,000 points are plotted in each diagram. The corresponding correlation coefficients $\rho$ are also denoted.}
\label{fig:comp}
\end{figure*}

In Fig. \ref{fig:comp} we compare our position angles $\phi^*$ to those by SDSS 
($\phi_\text{exp}$ and $\phi_\text{iso}$). The correlation of our position angle estimates with $\phi_\text{exp}$ is stronger than with $\phi_\text{iso}$, which is interesting as our method is essentially an isophote fitting method. The imprints of the strong bias in $\phi_\text{exp}$ are visible as horizontal stripes in the middle and right panes.

A drawback of our isophote fitting algorithm is that it does not account for seeing. As we do not deconvolve the PSF from the input images, smaller objects seem to be more circular. This is not expected to affect our final catalogue significantly because the median radius of our sample is 6'' in the $r$-band, much larger than the typical SDSS PSF of~1.4''~FWHM.


\section{Data and sample selection}
\label{sec:data}

We base our new catalogue of improved position angle measurements on $ u $, $ g $, $ r $, $ i $ and $ z $-band calibrated images taken from the SDSS~Stripe~82 data set \citep{Abazajian2009}. The Stripe~82 images were co-added from about $20$--$40$ individual exposures ad the reduced catalogue has got an estimated $ r $-band magnitude limit of about {$ r \lesssim 24 $~mag}, which is about one magnitude deeper than the rest of SDSS \citep{Gunn 1998}. The survey footprint is a $2.5^{\circ}$ wide stripe along the celestial equator ($-50^{\circ} \leq \alpha \leq 59^{\circ}$, $| \delta | \leq 1.26^{\circ}$).

The reason behind choosing the Stripe 82 data despite its limited footprint was that these co-added images provide far better signal-to-noise ratio than the single-exposure SDSS images, thus we can expect more reliable isophote fits.

\label{sec:sdss_selection}
We select objects from the co-added catalogue that are morphologically classified as galaxies and are brighter than a given magnitude in each band. Tab.~\ref{tab:obj_lst} lists the chosen magnitude limits for each photometric filter and the number of sources selected. Our algorithm outlined in Sec.~\ref{sec:algorithm} relies on precise source positions as input parameters; we take these positions from the Stripe~82 catalogue.

\section{Our unbiased catalogue}
\label{sec:cat}

Our image processing software outputs the fitted parameters with uncertainties and various other control parameters. To compile a catalogue from our data, we make use of the existing photometric and spectroscopic information from the SDSS Stripe 82 co-added survey. The catalogue contains the object identifiers, celestial coordinates, model magnitudes in each band, redshift, velocity dispersion, $eClass$ from the SDSS co-added catalogue, and, of course, the morphological parameters given by this work. These are the major axis, minor axis, the $X$ and $Y$ coordinates of the centroids and the position angle of the fitted ellipse in each photometric filter. Please note, that errors of the morphological parameters are likely to be overestimated, especially in case of the position angle, by the MPFIT routine. We include additional parameters: the number of edges of the contour polygons, the maximal intensity in a 3 by 3 pixel region around the ellipse centroids and around the original source coordinates.

\flushleft{The catalogue is available on-line at http://www.vo.elte.hu/galmorph.} 

\section{Discussion and future work}

The extreme bias identified in the position angle measurements of SDSS galaxies has strong implications in many fields of interest and might spoil important results. For instance, correlations between the orientation and spatial distribution of galaxies have long been studied but only the deep, large-scale optical surveys of the last decade made more elaborate investigation of this topic possible. Thus, it is extremely important to make sure these large catalogues contain reliable data. We are currently working on a more detailed statistical analysis of our algorithm that will be published in the near future. \\

\textit{Acknowledgements.} This work was supported by the following Hungarian grants: NKTH: Pol\'anyi, KCKHA005 and OTKA-103244.

\appendix

\end{document}